\documentclass[prd,twocolumn,floats,floatfix,nofootinbib,showkeys,showpacs]{revtex4-1}
\usepackage[dvips]{graphicx}
\usepackage{graphics}
\usepackage{dcolumn}
\usepackage{amssymb}
\usepackage{bm}
\usepackage{amsmath}
\usepackage{color}
\usepackage[utf8]{inputenc}

\def\spose#1{\hbox to 0pt{#1\hss}}

\def\lta{\mathrel{\spose{\lower 3pt\hbox{$\mathchar"218$}}
     \raise 2.0pt\hbox{$\mathchar"13C$}}}
\def\gta{\mathrel{\spose{\lower 3pt\hbox{$\mathchar"218$}}
     \raise 2.0pt\hbox{$\mathchar"13E$}}}
\newcommand{\be}{\begin{equation}}
\newcommand{\en}{\end{equation}}
\newcommand{\bea}{\begin{eqnarray}}
\newcommand{\ena}{\end{eqnarray}}

\graphicspath{{./figures/}}

\begin{document}

\title{On the Stability of
Cubic Galileon Accretion}

\author{Santiago P. E. Bergliaffa and Rodrigo Maier}

\email{rodrigo.maier@uerj.br}

\affiliation{Departamento de F\'isica Te\'orica, Instituto de F\'isica, Universidade do Estado do Rio de Janeiro,\\
Rua S\~ao Francisco Xavier 524, Maracan\~a, Rio de Janeiro, Brasil.}

\date{\today}

\begin{abstract}
We examine the stability of steady-state galileon accretion for the case of 
a Schwarzshild black hole. Considering the galileon action up to the cubic term
in a static and spherically symmetric background we obtain the general solution for the equation of motion
which is divided in two branches. By perturbing this solution we define an effective
metric which determines the propagation of fluctuations. In this general picture
we establish the
position of the sonic
horizon  together with the matching 
condition of the two branches on it.  
Restricting to the case of a Schwarzschild background, 
we show, via the analysis of the energy of the perturbations and its time derivative,
that the accreting field is linearly stable. 
\end{abstract}

\maketitle
\section{Introduction}

The description of the late-time accelerated expansion of the universe has motivated the study of many theories of modified gravity in the infrared limit (see \cite{Clifton2011} for a review). An example of such theories which displays interesting properties is 
the so-called galileon model
\cite{Nicolis2008} (see \cite{Deffayet2009} for a covariant generalization), which owes its name to an internal symmetry in which the gradient of the scalar field is 
shifted by a constant. Such a symmetry constrains the Lagrangian of the
theory 
to only five terms in 4 spacetime dimensions 
\cite{Nicolis2008}. 
The galileon model is 
the most general
scalar-tensor theory with equations of motion that contain no more than two derivatives (hence avoiding the Ostrogradski instability \cite{Ostrogradsky1850}).
Other important features \cite{Nicolis2008}
are (i) that it allows for the implementation of the Vainshtein mechanism
\cite{Vainshtein1972,
Babichev2013usa}, and (ii)
the non-quadratic kinetic coupling leads to the propagation of perturbations in an effective metric, as we shall see below \cite{Goulart2011}. 

The particular model that keeps only the first three terms of the Lagrangian of the galileon model, namely the cubic galileon, has been applied to the description of several phenomena, such as 
compact objects  \cite{Chagoya2014} and black holes in a cosmological setting \cite{Babichev2016}.
Limits on the coupling constants of this model coming from terrestrial experiments have been obtained in \cite{Brax2011}, and from cosmological observations in \cite{Neveu2016}.
Its covariant version has been used to study the large-scale stucture problem \cite{Bhattacharya2015}
and tested with the Coma Cluster \cite{Terukina2015}. 
As with every new theory, 
it is important to continue the examination of the consequences of the cubic galileon, in particular by checking the existence and stability of relevant solutions. This task that was initiated in \cite{Babichev2010}, where the steady-state and 
spherically symmetric accretion of a galileon field onto a Schwarzchild black hole in the test fluid approximation was analyzed (both for the cubic galileon, and for the combination of the second and fourth terms of the galileon model). 
Specifically, the conditions for the existence 
of the critical flow were established, 
as well as the dependence of the position of the sonic horizon with the parameters of the theory.
Here we shall tackle the problem of the linear stability of the accretion of the cubic galileon analyzed in \cite{Babichev2010}, using a method developed by Moncrief
\cite{Moncrief1980}. Such a method
is based 
on the evaluation of the sign of the time derivative of the energy of the perturbations in a given volume \footnote{This method was used in \cite{Paz2014} to establish the linear stability of the accretion 
of a ghost condensate onto a Schwarzschild black hole, 
and in 
\cite{Mach2013} to analyze Bondi accretion in Schwarzschild-(anti-)de Sitter spacetimes.}. A negative sign together with the positivity of the energy of the perturbations implies 
linear stability. 
By use of Moncrief's method, we shall show that the abovementioned system is linearly stable.

\section{Preliminary Setting}

\par
Let us consider a test galileon field $\phi$ whose action reads
\begin{eqnarray}
\label{eq1}
S_\phi=\int \sqrt{-g}\; [\alpha \nabla_\mu\phi\nabla^\mu\phi + \beta (\nabla_\mu\phi\nabla^\mu\phi)\square\phi] d^4x,
\end{eqnarray}
where $\alpha>0$ and $\beta$ are coupling constants and 
$g$ is the determinant of the background metric. 
It is worth mentioning that here we do not take into account the ``potential'' 
term proportional to $\phi$.
In fact, as shown in \cite{Babichev2010} there is no steady-state solution for the
accretion once such term is considered.

Varying the action (\ref{eq1}) with respect to the scalar field $\phi$ 
and adopting the following convention 
\begin{eqnarray}
\label{eq13b}
\nabla_\mu\nabla_\nu\nabla_\beta\phi-\nabla_\nu\nabla_\mu\nabla_\beta\phi=-R^\sigma_{~\beta\mu\nu}\nabla_\sigma\phi
\end{eqnarray}
for the Riemann tensor, we obtain
that the equation of motion is given by
\begin{eqnarray}
\label{eq14n}
\nonumber
\square\phi+\gamma[(\square\phi)^2 - R_{\mu\nu}(\nabla^\mu\phi)(\nabla^\nu\phi)~~~~~~~~~~~~~~~~
\\
- (\nabla^\mu \nabla^{\nu} \phi) (\nabla_\mu \nabla_{\nu} \phi)]=0,
\end{eqnarray}
where $\gamma\equiv\beta/\alpha$.
Defining
\begin{eqnarray}
\label{eq14new}
j^\mu=2\nabla^\mu\phi+\gamma[2\nabla^\mu\phi\square\phi-\nabla^\mu(\nabla_\nu\phi\nabla^\nu\phi)],
\end{eqnarray}
the above equation of motion may also be written as 
\begin{eqnarray}
\label{eq14newer}
\nabla_\mu j^\mu=0.
\end{eqnarray}

The variation of Eqn.(\ref{eq1}) with respect to the metric $g^{\mu\nu}$
yields the following energy-momentum tensor:
\begin{eqnarray}
\label{eqtmunu}
\nonumber
T_{\mu\nu}=\alpha (2\nabla_\mu\phi\nabla_\nu\phi-g_{\mu\nu}\nabla_\mu\phi\nabla^\mu\phi)~~~~~~~~~~~~~~~~
\\
+ \beta [2\square\phi\nabla_\mu\phi\nabla_\nu\phi-\nabla_\mu\phi\nabla_\nu(\nabla_\mu\phi
\nabla^\mu\phi)].
\end{eqnarray}
We shall show next that the dynamics of the perturbations of the cubic galileon field is governed by an effective metric
\footnote{See \cite{Barcelo2005} for a review of the effective metric, and \cite{Goulart2011} for the case of a scalar field.}. As discussed below, the effective metric is related to the liner stability of the system.
Let us perturb the background field solution in such a way that 
\begin{eqnarray*}
\phi(t,r,\theta , \varphi)=\phi_0(t,r)+\phi_1 (t,r,\theta , \varphi).
\end{eqnarray*}
Feeding this expression in the equation of motion (\ref{eq14n}) , we obtain
\begin{eqnarray}
\label{eqb0}
\nonumber
\partial_\nu\{\sqrt{-g}[(1+2\gamma\square\phi_0)g^{\mu\nu}-2\gamma\nabla^\mu\nabla^\nu\phi_0]\nabla_\nu\phi_1\}=0.
\end{eqnarray}
Defining the effective metric $\tilde{g}^{\mu\nu}$ as
\begin{eqnarray}
\label{eqc}
\sqrt{-\tilde{g}}\tilde{g}^{\mu\nu}=\sqrt{-g}[(1+2\gamma\square\phi_0)g^{\mu\nu}-2\gamma\nabla^\mu\nabla^\nu\phi_0],~~~
\end{eqnarray}
the equation of motion for the perturbations reduces to
\begin{eqnarray}
\label{eqb1}
\partial_\nu\{\sqrt{-\tilde{g}}\tilde{g}^{\mu\nu}\tilde{\nabla}_\nu\phi_1\}=0,
\end{eqnarray}
where $\tilde{\nabla}_\nu$ is the covariant derivative built with 
the effective metric. 
It follows from Eqn.(\ref{eqc}) that the effective metric can be written as
\begin{eqnarray}
\label{eqc1}
\tilde{g}^{\mu\nu}=\frac{M^{\mu\nu}}{\sqrt{-g}\sqrt{-M}}
\end{eqnarray}
where $M^{\mu\nu}=(1+2\gamma\square\phi_0)g^{\mu\nu}-2\gamma\nabla^\mu\nabla^\nu\phi_0$
and $M={\rm det}(M^{\mu\nu})$. 
Therefore, if the factor $(\sqrt{-g}\sqrt{-M})^{-1}$ is regular (and we shall see below that this is the case in
the problem analyzed here), 
$M^{\mu\nu}$ can be taken as the effective metric tensor.

We shall set next the stability conditions, as discussed in \cite{Moncrief1980}. 
Since the perturbations obey the equation of motion of a massless scalar field in the effective metric (see Eqn.(\ref{eqb1})), it follows that 
the associated action for the perturbations is given by
\begin{eqnarray}
\label{eqd}
S=\int\sqrt{-\tilde{g}}\;\tilde{g}^{\alpha\beta}\tilde{\nabla}_\alpha\phi_1\tilde{\nabla}_\beta\phi_1d^4x.
\end{eqnarray}
The variation of this action with respect to $\tilde g^{\mu\nu}$ furnishes the
energy-momentum tensor for $\phi_1$, given by
\begin{eqnarray}
\label{eqd1}
\tilde{T}_{\mu\nu}=\tilde{\nabla}_\mu\phi_1\tilde{\nabla}_\nu\phi_1-\frac{1}{2}\tilde{g}_{\mu\nu}\tilde{\nabla}_\alpha\phi_1\tilde{\nabla}^\alpha\phi_1.
\end{eqnarray}
The conservation equation $\tilde{\nabla}_{\mu}\tilde{T}^{\mu}_{~\nu}=0$ is automatically
satisfied taking into account the equation of motion for
the
perturbations. 

As shown in \cite{Moncrief1980}, if $Z^\mu$ is a Killing vector of the
effective metric, it follows that
$\tilde{\nabla}_\mu(Z^\nu \tilde{T}^\mu_{~\nu})=0$, which  can be written as
\begin{eqnarray}
\label{eqe}
  \partial_\mu(\sqrt{-\tilde{g}}Z^\nu \tilde{T}^\mu_{~\nu})=0.
\end{eqnarray}
Choosing $Z_\mu=\delta^{t}_{~\mu}$, integrating the above
equation in a $3$-volume $V$,
and using Gauss's theorem, we obtain
\begin{eqnarray*}
\frac{d}{dt}\int_V(\sqrt{-\tilde{g}}\tilde{T}^t_{~t})d^3x=-
\int_\Sigma(\sqrt{-\tilde{g}}\tilde{T}^i_{~t})d\Sigma_i ,
\end{eqnarray*}
where $\Sigma$ is the surface enclosing the volume $V$.
Using the definition of the  energy of the perturbations, given by
\begin{eqnarray}
\label{eqen}
\tilde{E}=\int_V(\sqrt{-\tilde{g}}\tilde{T}^t_{~t})d^3x
\end{eqnarray}
it follows that  \cite{Moncrief1980}
\begin{eqnarray}
\label{eqh}
\frac{d\tilde{E}}{dt}=-
\int_\Sigma(\sqrt{-\tilde{g}}\tilde{T}^i_{~t})d\Sigma_i.
\end{eqnarray}
An appropriate choice of the surface $\Sigma$ will allow the determination of the sign of the RHS of Eqn.(\ref{eqh}) without actually carrying out the integration. Such a sign, together with the finiteness and positivity of the energy of the perturbations, determines whether the system is linearly stable or not.

\section{The Model}

In what follows we shall consider a background geometry given by 
\begin{equation}
\label{eq11}
ds^2=F(r)dt^2-\frac{1}{G(r)}dr^2-r^2(d\theta^2+\sin^2{\theta}d\varphi^2).
\end{equation}
The steady-state assumption entails that the background field $\phi_0$ takes the form
\begin{equation}
\label{eq12}
\phi_0(t,r)=t+\psi(r).
\end{equation}
Using these equations in Eqn.(\ref{eq14newer}) the following first integral
for the equation of motion (\ref{eq14n}) is obtained:
\begin{eqnarray}
\label{eqfi0}
r^2\sqrt{\frac{F}{G}}j^r= B
\end{eqnarray}
where $B$ is a positive arbitrary constant,
\begin{eqnarray}
\label{eqfi1}
j^r=-2 G \psi^\prime - \gamma G \Big[\frac{F^\prime}{F^2}
-G{\psi^\prime}^2\Big(\frac{F^\prime}{F}+\frac{4}{r}\Big)\Big],
\end{eqnarray}
and the prime denotes the derivative with respect to $r$. It follows from Eqn.(\ref{eqfi1}) that the derivative of
the general solution
of the equation of motion (\ref{eq14n}) is given by
\begin{eqnarray}
\label{eq13}
\nonumber
\psi_{(\pm)}^\prime(r)=\frac{1}{\gamma FG^2r(4F+rF^\prime)}\Big\{GF^2r^2\pm\sqrt{rFG^2}\times
\\
\Big[ r^3F^3+\gamma (4F+rF^\prime)\Big(BF^2\sqrt{\frac{G}{F}}+\gamma r^2GF^\prime\Big)\Big]^{1/2}
\Big\}.
\end{eqnarray}
Using the definitions
$v(r)\equiv F(r)\psi^\prime(r)
%
$ and $H(r)\equiv\sqrt{-g}\sqrt{-M}$,
the components of the effective geometry 
are given by
\begin{eqnarray}
\label{eqeg}
\nonumber
\tilde{g}^{tt}& = &\frac{1}{HF}\Big\{1-\gamma\Big[v\Big(\frac{4G}{rF}+\frac{2GF^\prime}{F^2}-\frac{G^\prime}{F}\Big)-\frac{2Gv^\prime}{F}\Big]\Big\}
\\
\nonumber
\tilde{g}^{tr}& = &-\frac{\gamma  G F^\prime}{HF^2}\\
\nonumber
\tilde{g}^{rr}& = &-\frac{G}{H}\Big[ 1- \frac{\gamma G v}{rF^2}(4F+rF^\prime)\Big]\\
\nonumber
\tilde{g}^{\theta\theta}&=&-\frac{1}{Hr^2}\Big\{1+\gamma\Big[v\Big(\frac{GF^{\prime}}{F^2}-\frac{G^\prime}{F}-\frac{2G}{rF}\Big)-\frac{2Gv^\prime}{F}\Big]\Big\}
\\
\nonumber
\tilde{g}^{\varphi\varphi}& = &\tilde{g}^{\theta\theta}\sin^{-2}{\theta}
\end{eqnarray}
The position of the putative sound horizon for the perturbations (denoted $r_s$) must 
satisfy the condition $\tilde{g}^{rr}\Big|_{r_s}=0$, which entails
\begin{equation}
\label{eqrs}
 v\Big|_{r_s}=\frac{r F^2}{\gamma G(4F+rF^\prime)}\Big|_{r_s}.
\end{equation}
This expression will be used below in the equation of motion for the background to get
the radius of the sonic horizon(s).
The first integral displayed in Eqn.(\ref{eqfi0}) may also be written as
\begin{eqnarray}
\label{eqfi2}
\nonumber
\frac{2 G v}{F} + \gamma G \Big[\frac{F^\prime}{F^2}
-G\frac{v^2}{F^2}\Big(\frac{F^\prime}{F}+\frac{4}{r}\Big)\Big]~~~~~~\\
+\sqrt{\frac{G}{F}}\frac{B}{r^2}=0.
\end{eqnarray}
Differentiating this equation with respect to $r$, 
we obtain the equation of motion (\ref{eq14n}) in the form
\begin{eqnarray}
\label{eqh1}
\nonumber
4r^2F^2M^{rr}v^\prime=
4r^2FM^{rr}vF^\prime~~~~~~~~~~~~~~~~~~~~~~~~~~~~~~~
\\
\nonumber
+2 F v(4FG+rGF^\prime+rFG^\prime)~~~~~~~~~~~~~~~~~~~~\\
\nonumber
+\gamma\Big\{ r [F F^\prime (4G+rG^\prime) + 2rGFF^{\prime\prime} - 3rG(F^\prime)^2 ]\\
\nonumber
+\frac{v^2}{F^2}[r^2G^2F(F^\prime)^2-rGF^2F^\prime(8G+3rG^\prime)\\
-2F^2G(4FG+6rFG^\prime+r^2GF^{\prime\prime})]\Big\}.~~~~~~~~~
\end{eqnarray}
Evaluating the above equation at the sound horizon and
using Eqn.(\ref{eqrs}) 
we end up with the requirement
\begin{eqnarray}
\label{eqh2}
\nonumber
\Big\{-r[F^\prime(4FG-3rGF^\prime+rFG^\prime)+2rGFF^{\prime\prime}]\\
\nonumber
+\frac{ r^2 F^3}{\gamma^2 G(4F+rF^\prime)^2}
\Big[r[2F(2FG^\prime+rGF^{\prime\prime})\\+FF^\prime(rG^\prime-8G)-3rG(F^\prime)^2]-24F^2G\Big]\Big\}\Big|_{r_s}=0,~~~~
\end{eqnarray}
in order to ensure that $v^{\prime}$ is regular at $r_s$. 
The roots of Eqn.(\ref{eqh2}) 
give the position of the 
sound horizons for  
a cubic galileon accreting onto a 
spherically symmetric
black hole.

In what follows we shall choose the functions $F$ and $G$
as those of a Schwarzschild black hole
in rescaled units, namely, 
\begin{eqnarray}
\label{eqh3}
F(r)=G(r)=1-\frac{1}{r}.
\end{eqnarray}
In this case we obtain 
\begin{eqnarray}
\label{eqh5}
\tilde{g}^{rr}
=\frac{1}{H}\{-F+\gamma v [1+(2-3F)F]\},
\end{eqnarray}
so that the horizon condition is given by
\begin{eqnarray}
\label{eqh6}
v|_{r_s} = \frac{ F}{\gamma [1 + (2 - 3 F) F]}\Big|_{r_s},
\end{eqnarray}
which implies that $\tilde{g}_{tt}(r_s)\equiv 0$.\\
In terms of $v$, the equation of motion reads
\begin{eqnarray}
\label{eqh7}
\nonumber
2FM^{rr}v^\prime=(1-F)\{4 F^2 v~~~~~~~~~~~~~~~~~~~~~~~~\\
+ \gamma [(-1 + F)^3 + (F^2-1) (3 F-1) v^2]\},
\end{eqnarray}
so that the requirement (\ref{eqh2}) turns into
\begin{eqnarray}
\label{eqh8}
\nonumber
\{F^2[1+F(2+9F)]~~~~~~~~~~~~~~~~~~~~~~~~~~~~\\
-\gamma^2(F-1)^4 (1+3F)^2\}|_{r_s}=0.
\end{eqnarray}
For $r\rightarrow \infty$, we obtain from Eqn.(\ref{eq13}) that
\begin{eqnarray}
\label{eqh8h1}
\psi^\prime_{(+)}(r)\sim \frac{r}{2\gamma}
\end{eqnarray}
and
\begin{eqnarray}
\label{eqh8h2}
\psi^\prime_{(-)}(r)\sim -\frac{1}{2 r^2}(B+\gamma)
-\frac{1}{2 r^3}(B+2\gamma).
\end{eqnarray}
While the result given in Eqn.(\ref{eqh8h1}) violates the assumption of homogeneity of
the solution at spatial infinity, the solution given by Eqn.(\ref{eqh8h2})
is well-behaved in that limit.
As pointed out in \cite{Babichev2010}, 
the value of $B$ must be set by imposing that  $\psi_{+}$ and $\psi_{-}$ and their derivatives match
at $r=r_s$. 
The surface $r =r_s$ (also called the critical point)
is the acoustic horizon in the sense that the perturbations 
$\phi_1$ inside this surface cannot  escape towards 
the asymptotically flat region. For $\gamma\neq 0$,  condition (\ref{eqh8}) yields
only one real root $r_s>1$. Moreover, in order to obtain the matching
of $\psi_{\pm}|_{r_s}$ and their derivatives for $r_s>1$, $\gamma$ must be negative.
From now on we are going to restrict ourselves to configurations
such $\gamma<0$.

\begin{figure}
\includegraphics[width=8cm,height=6cm]{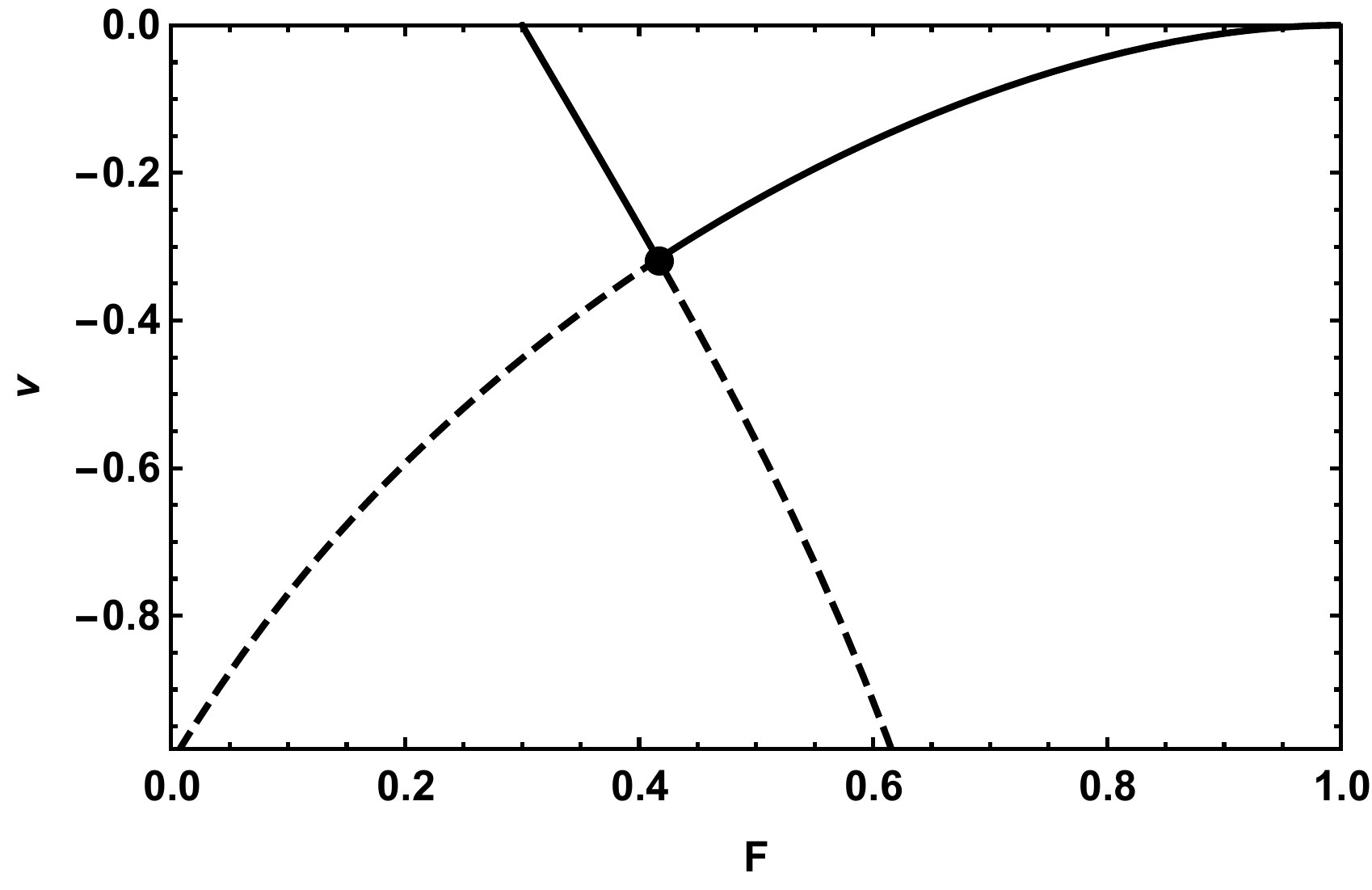}
\caption{Numerical plot of the first integral given in Eqn.(\ref{eqh9}).
The dashed (solid) line refers to $\psi^\prime_{(+)}$, 
($\psi^\prime_{(-)})$. Both solutions match at $r=r_s$ as
expected. Here we choose $\gamma=-1$ so that $r_s=1.71286$
($F|_{r_s} = 0.41618$) and $B=3.33294$.}
\label{firstint}
\end{figure}

For the Schwarzschild case, the first integral of the equation of motion, Eq.(\ref{eqh7}),
is given by
\begin{eqnarray}
\label{eqh9}
\nonumber
v^2\gamma(3F+1)(F-1)~~~~~~~~~~~~~~~~~~~~~~~~~~~~~~~~~~\\
+2 Fv+(F-1)^2( \gamma + B F)=0,
\end{eqnarray}
so that
\begin{eqnarray}
\label{eqh91}
v_{(\pm)}=\frac{F(1\pm \sqrt{1-\Delta})}{\gamma(1+3F)(1-F)},
\end{eqnarray}
where
\begin{eqnarray}
\label{eqdel}
\Delta\equiv{\gamma F^{-2}(F-1)^3(1+3F)(BF+\gamma)}.
\end{eqnarray}
As an illustration, let us take $\gamma=-1$. The sound horizon (for $r_s > 1$) 
is located at $r_s=1.71286$
(with $F|_{r_s} = 0.41618$).  
The matching of the solutions at $r_s$ fixes the value of the constant $B$, given by  $B=3.33294$. A
numerical plot of the
first integral (\ref{eqh9}) for this case is shown in Fig.\ref{firstint}.

\begin{figure}
\includegraphics[width=8cm,height=6cm]{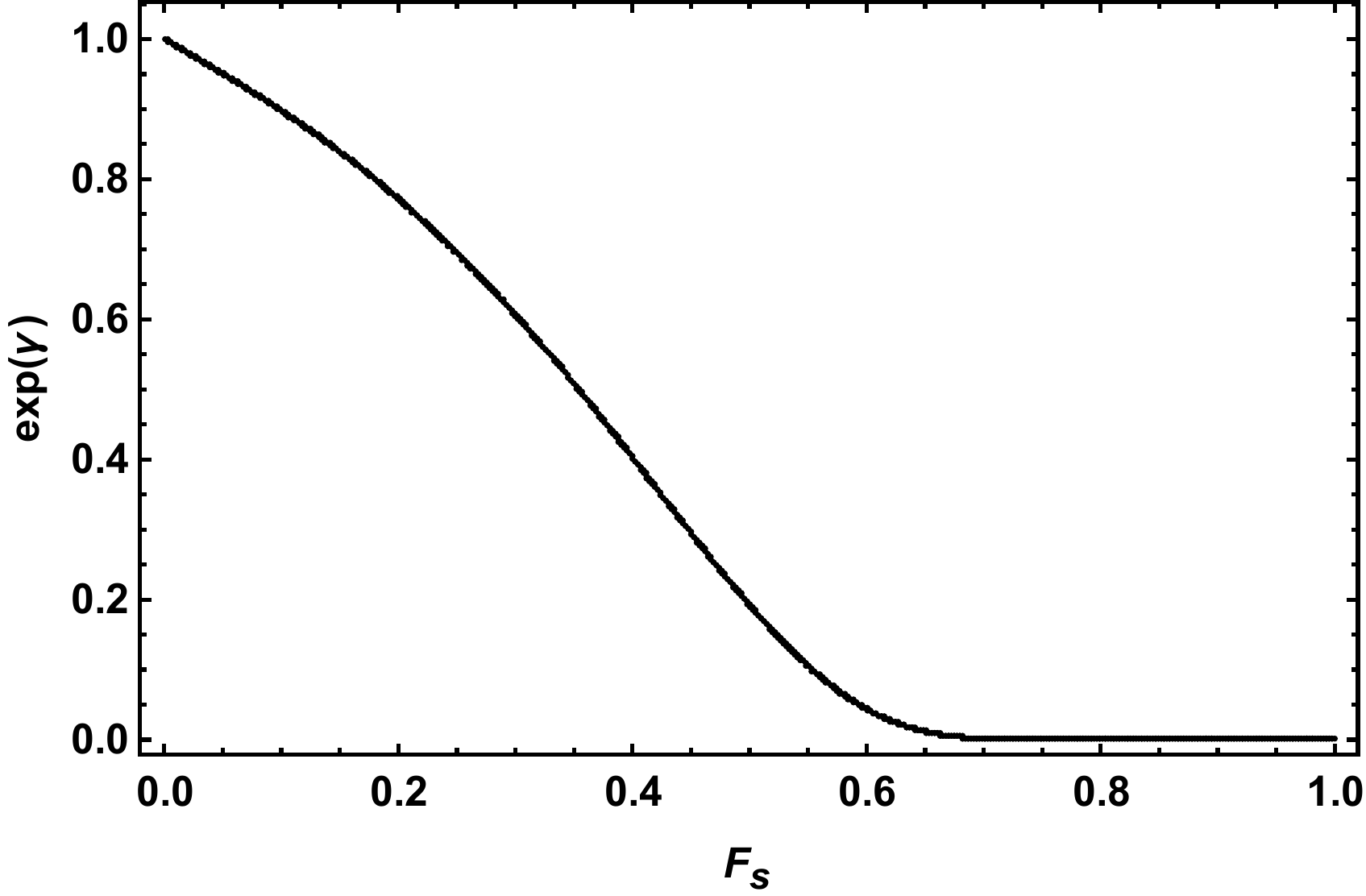}
\caption{Numerical plot of $\exp{(\gamma)}$
as a function of $F_s$. Here we see that the position of the sound horizons
$F_s$ moves towards
the Schwarzschild radius as $|\gamma|\rightarrow 0$.
On the other hand, $F_s\rightarrow 1$ as $|\gamma|\rightarrow +\infty$.
}
\label{sonhor}
\end{figure}

It is worth noting that as $|\gamma|$ increases, the position of the
sonic horizon moves further away from the event horizon. We illustrate
this behaviour through the plot built by numerical evaluation of Eqn.(\ref{eqh8}), shown in Fig. \ref{sonhor}. 

\section{Stability}

Let us now go back to the evaluation of the stability of the system. A convenient choice for the volume $V$ is that encompassed between the
surfaces $r=r_s$ and $r\rightarrow \infty$. 
We shall decompose accordingly the integral in Eqn.(\ref{eqh}), 
$$
\frac{d\tilde E}{dt} = I_{r_s} + I_\infty.
$$
It follows from 
Eqns.(\ref{eqc}) 
and (\ref{eqh8h2}) that, 
for $r\rightarrow \infty$,
\begin{eqnarray}
\label{eqs1}
\sqrt{-\tilde{g}}\tilde{g}^{rr}\rightarrow \sqrt{-{g}}{g}^{rr}=- r^2\sin{\theta}
\end{eqnarray}
and $\tilde{g}^{tr}\rightarrow 0$. Therefore, 
\begin{eqnarray}
\label{eqs2}
I_\infty = -\int_{\Sigma_\infty}r^2\sin\theta(\partial_t\phi_1\partial_r\phi_1)|_{r\rightarrow\infty}d\Sigma_{r}
\end{eqnarray}
The assumption that the perturbations fall off at least as
$1/r^p$ (with $p>0$) together with the finiteness of the energy
of the perturbations (see Eqn.(\ref{eqen})), 
implies that 
\begin{eqnarray}
\label{eqs4}
\partial_t \phi_{1}\sim\frac{1}{r^{\frac{3}{2}+\epsilon}},~\partial_r \phi_{1}\sim\frac{1}{r^{\frac{5}{2}+\epsilon}},~\epsilon>0,
\end{eqnarray}
so that $I_\infty$  vanishes. Hence, the integral on the l.h.s. of Eqn.(\ref{eqh}) reduces to
\begin{eqnarray}
\label{eqs5}
\frac{d\tilde{E}}{dt}=-\int_{\Sigma_{r_s}}[\sqrt{-\tilde{g}}(\partial_t \phi_{1})^2\tilde{g}^{rt}]|_{r_s}d\Sigma_{r_s}.
\end{eqnarray}
From Eq.(\ref{eqc}),
we have that $\sqrt{-\tilde{g}}\tilde{g}^{tr}=-2\gamma\sqrt{-g}\nabla^t\nabla^r\phi_0$ where 
\begin{eqnarray}
\label{eqs6}
\nabla^t\nabla^r\phi_0=\frac{1}{2r(r-1)}.
\end{eqnarray}
Since $\gamma<0$ and $r_s>1$, it follows  that the energy of the perturbations decreases with
time. 

To conclude that the system is stable, we still need to
show that the energy of the perturbations
is positive definite. It follows from Eqns.(\ref{eqd1}) and (\ref{eqen}) that
\begin{eqnarray}
\label{eqs7}
\tilde E =  \int_V \sqrt{-\tilde g}\; [\tilde{g}^{tt}(\partial_t\phi_1)^2
-\tilde{g}^{ii}(\partial_i\phi_1)^2 ] .
\end{eqnarray}
It suffices then to
show that 
$\tilde{g}^{tt}>0$ and $\tilde{g}^{ii}<0$ for $r>r_s$, with $i=r,\theta , \varphi$.

\begin{figure}
\includegraphics[width=6cm,height=6cm]{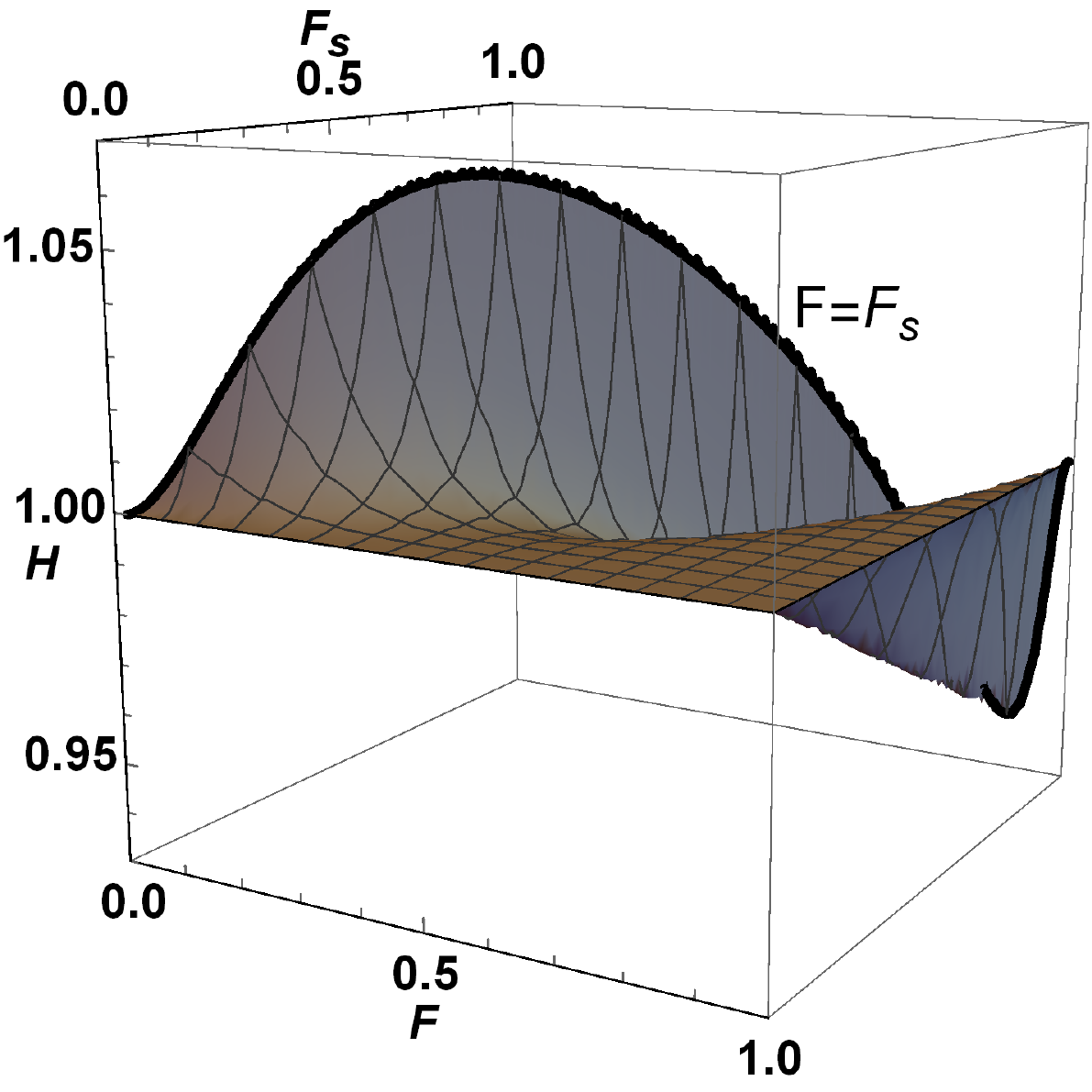}
\caption{The function $H(r)\equiv \sqrt{-g} \sqrt{-M}$ is ploted 
in terms of
$F$ and $F_s$. The solid black line defines the contour $F=F_s$.}
\label{detM}
\end{figure}
\begin{figure}
\includegraphics[width=6cm,height=6cm]{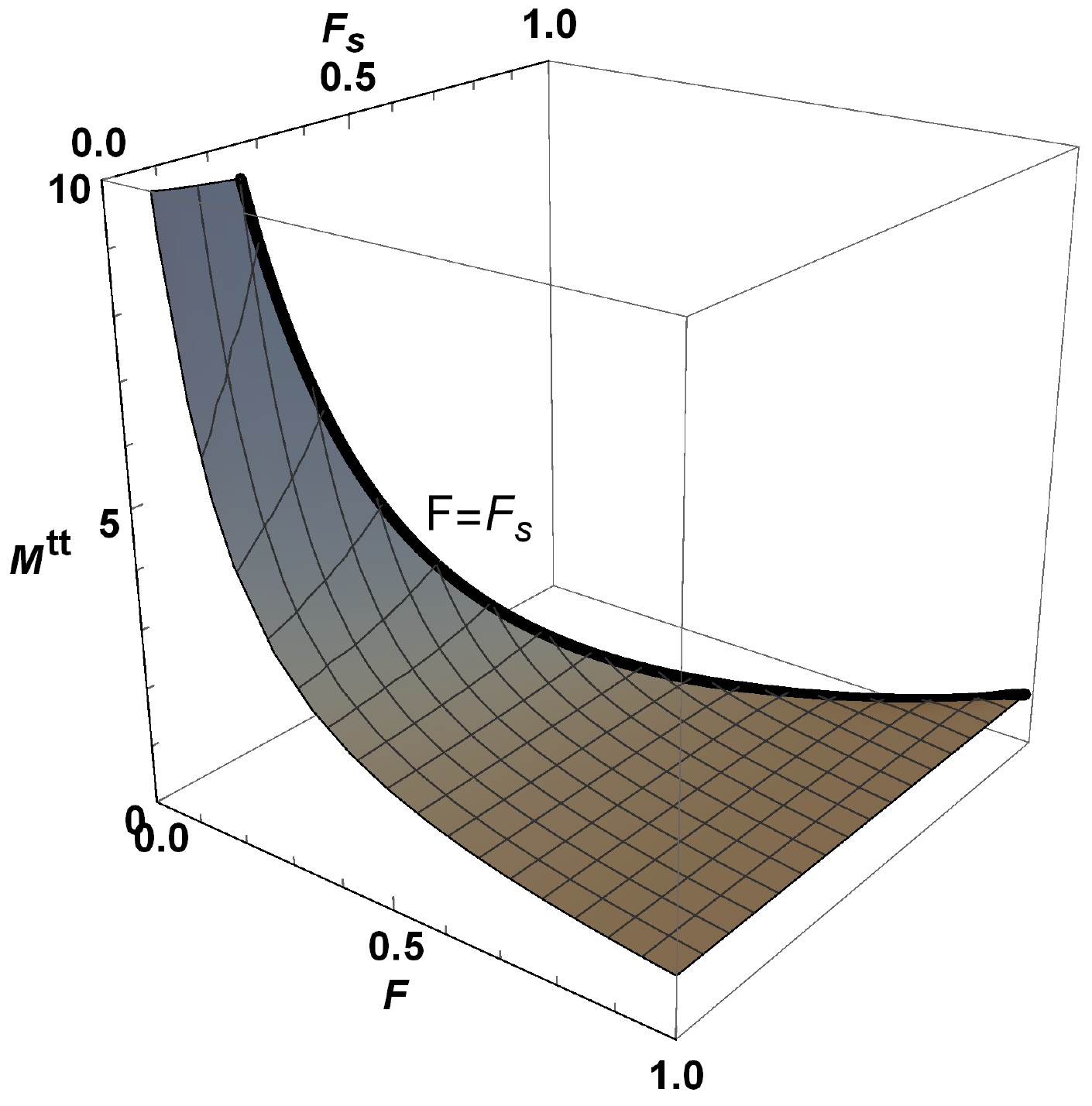}
\caption{The metric coefficient $M^{tt}$ as a function of $F$ and $F_s$ for $F_s<F<1$. The solid black line defines the contour $F=F_s$.}
\label{mttn}
\end{figure}
\begin{figure}
\includegraphics[width=6cm,height=6cm]{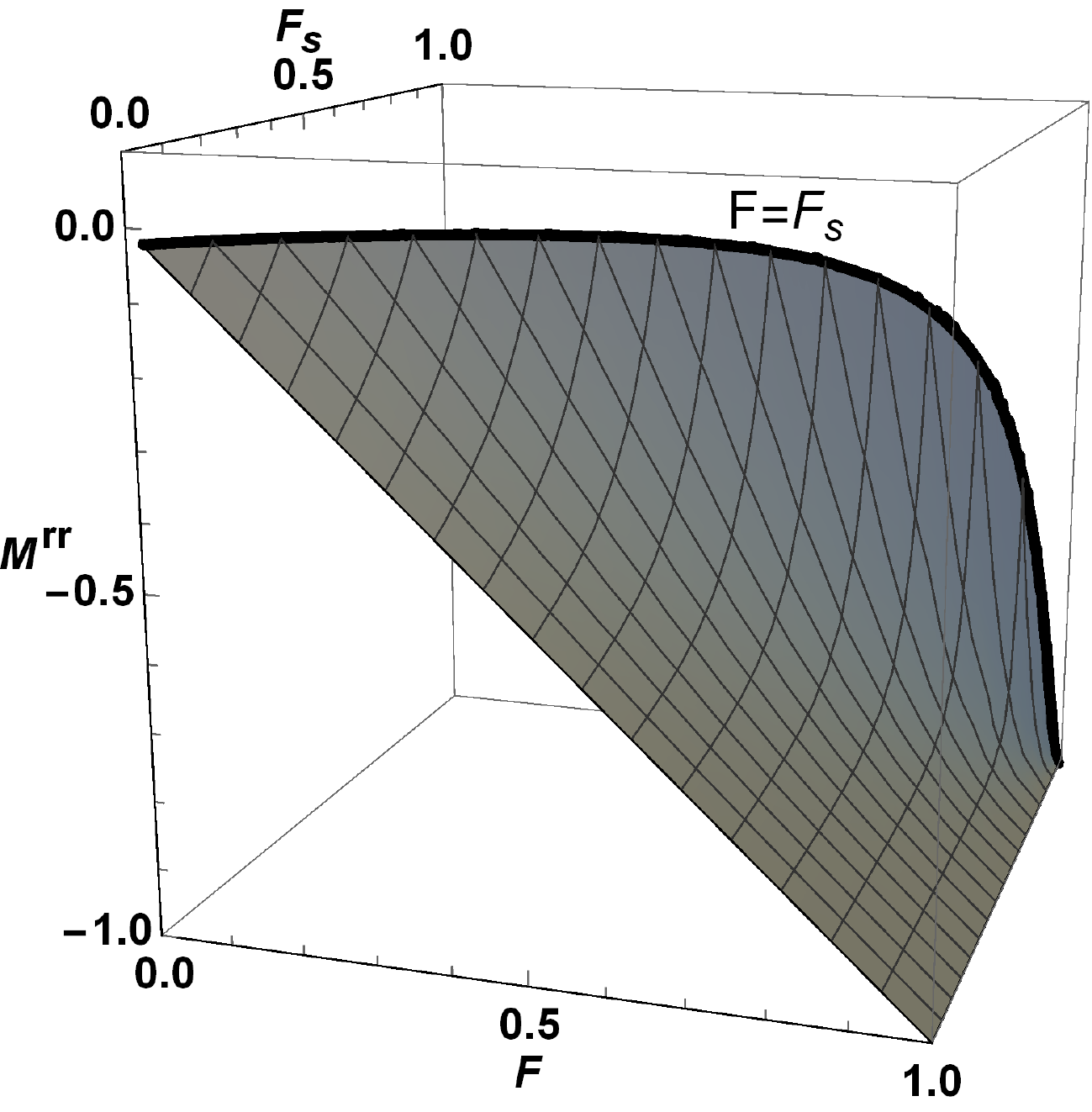}
\caption{The metric coefficient $M^{rr}$ as a function of $F$ and $F_s$ for $F_s<F<1$. The solid black line defines the contour $F=F_s$.}
\label{Mrrn}
\end{figure}
\begin{figure}
\includegraphics[width=6cm,height=6cm]{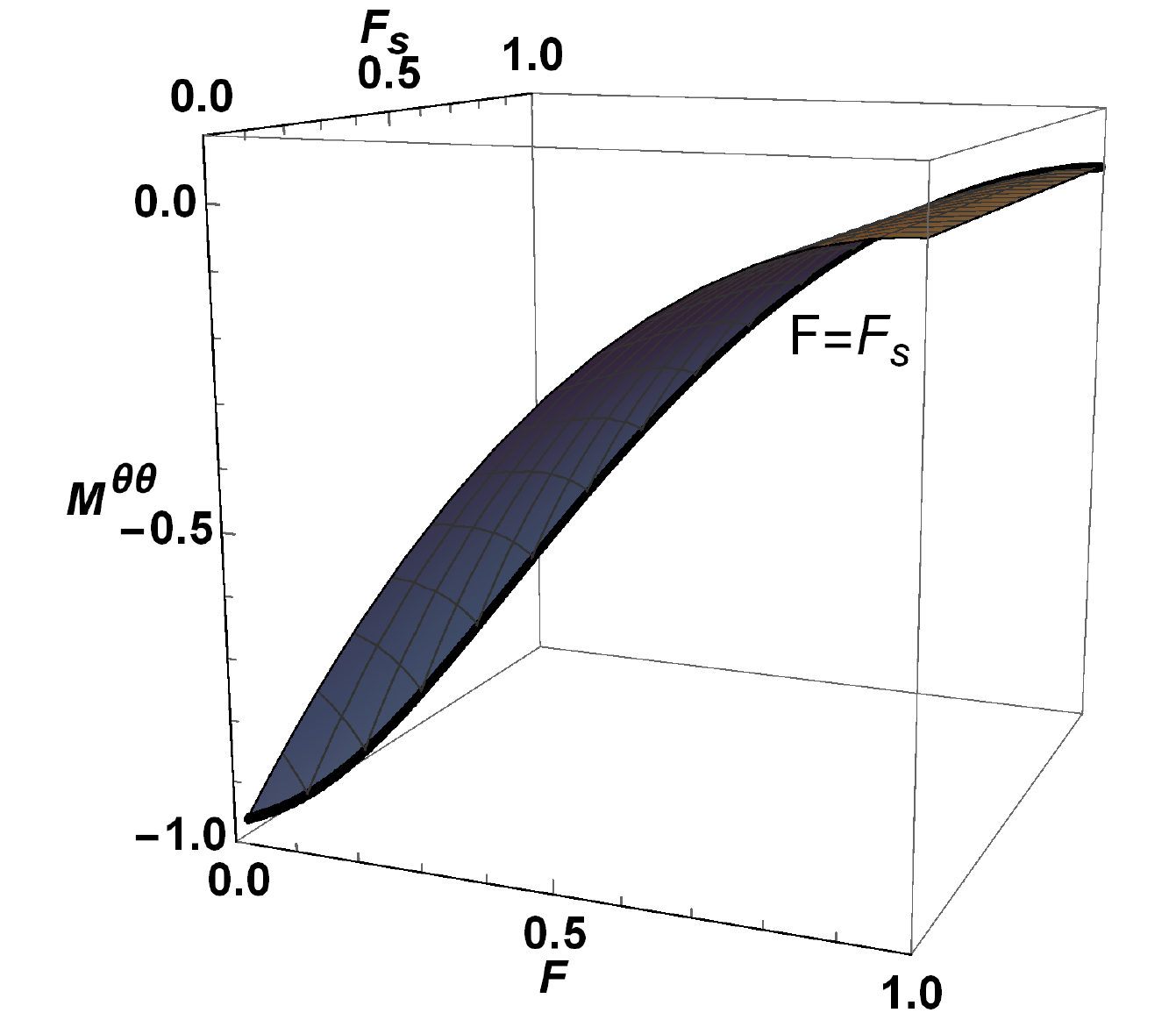}
\caption{The metric coefficient $M^{\theta\theta}$ as a function of $F$ and $F_s$ for $F_s<F<1$. The solid black line defines the contour $F=F_s$.}
\label{Mthetathetan}
\end{figure}

Since the behaviour of 
$H(r)=\sqrt{-g}\sqrt{-M}$ is shown to be regular for all values of the relevant variables in
Fig. \ref{detM} (see below for details of this and the next plots),
from now on we will take  $M^{\mu\nu}$ as the effective metric. For the Schwarzschild case 
we obtain
\begin{eqnarray}
\label{eqMmunu}
M^{tt}&=&\frac{F+\gamma [(1-6F+5F^2)v-2Fv^\prime]}{F^2},\\
M^{rr}&=&\gamma v-F[1+\gamma v(3F-2)],\\
\label{eqMmunu3}
M^{\theta\theta}&=&-(F-1)^2[1+2\gamma v (F-1)-2\gamma v^\prime].
\end{eqnarray}
These expressions are a function of $r$ (through $F$, $v$ and $v'$), and depend on the parameters $\gamma$ and $B$ (through $v$ and $v'$). They can be rewritten in terms of $F_s$ and $F$ as follows
\footnote{All the expressions in the following calculations are very lengthy, so we shall outline the procedure we followed, and plot the resulting equations.}.
The constant $B$ can be expressed in terms of $\gamma$ and $F_s$ using the matching condition 
$(\psi_+ - \psi_-)|_{F_s}=0$. 
Now $\gamma$ can be written in terms of $F_s$ by use of the equation (\ref{eqh8}) evaluated at $F_s$, which yields a second order polinomial in
$\gamma$. We choose to work with the negative root since 
$\gamma$ must be negative, as implied by the matching conditions (see Sect. III).
We have then $\gamma = \gamma (F_s)$, and $B=B(F_s,\gamma(F_s))$. These two relations allow us to 
write the metric coefficients given above as functions of $F$ and $F_s$ only.
We present in Figs. \ref{mttn}, \ref{Mrrn}, and \ref{Mthetathetan} the plots of the metric coefficients for $F_s<F<1$. The solid black lines define the contours $F=F_s$. 
The plots show that each coefficient has the sign required to make the energy of the perturbations positive
for any value of $F$ and $F_s$ in the relevant
interval (that is, for any value of $\gamma <0$). Hence, we conclude that the system is linearly stable. 

\section{Conclusions}

We have shown that the system composed of a
Schwarzschild black hole accreting a steady-state and
spherically symmetric galileon field described by the action given in Eqn.(\ref{eq1}) 
is linearly stable using the method developed by Moncrief \cite{Moncrief1980}. This method rests in the 
determination of the sign of the time derivative of the energy of
the perturbations through a surface integral and leads in a
few steps to the linear stability of any stationary solution of
the system at hand that goes from being homogeneous at
infinity, passes through a sonic horizon, and reaches the
Schwarzschild horizon. The method profits from the
symmetries of the system, and it does not use the explicit
form of the solution for the scalar field. Extensions of this work, currently
under way, are the study of the nonlinear stability of the
system (since linear stability is only a prerequisite for full
stability), and 
the determination of 
the stability of the 
accretion in the case of the  model described by 
the combination of the first,  second and fourth terms of the galileon Lagrangian.

\section*{Acknowledgements}

S.E.P.B. would like to acknowledge support from
Fundação de Amparo à Pesquisa do Estado do Rio de
Janeiro (FAPERJ) and Universidade do Estado do Rio de
Janeiro (UERJ).
Figures were generated
using the Wolfram Mathematica $7$.

\end{document}